\documentclass[12pt]{article}
\usepackage[colorlinks,allcolors=blue,bookmarks=false,hypertexnames=true]{hyperref}
\usepackage{fullpage,graphicx,psfrag,amsmath,amsfonts,verbatim}
\usepackage{url}
\usepackage[small,bf]{caption}

\newcommand{\ones}{\mathbf 1}
\newcommand{\reals}{{\mbox{\bf R}}}

\newcommand{\symm}{{\mbox{\bf S}}}  

\newcommand{\Tr}{\mathop{\bf Tr}}

\newcommand{\eg}{{\it e.g.}}
\newcommand{\ie}{{\it i.e.}}

\newcommand{\BEAS}{\begin{eqnarray*}}
\newcommand{\EEAS}{\end{eqnarray*}}
\newcommand{\BEA}{\begin{eqnarray}}
\newcommand{\EEA}{\end{eqnarray}}
\newcommand{\BEQ}{\begin{equation}}
\newcommand{\EEQ}{\end{equation}}
\newcommand{\BIT}{\begin{itemize}}
\newcommand{\EIT}{\end{itemize}}

\title{Portfolio Construction Using Stratified Models}
\author{Jonathan Tuck \and Shane Barratt \and Stephen Boyd}

\begin{document}
\maketitle

\begin{abstract}
In this paper we develop models of asset return mean and covariance
that depend on some observable market conditions, and use these
to construct a trading policy that depends on these conditions,
and the current portfolio holdings.
After discretizing the market conditions,
we fit Laplacian regularized stratified models for the return mean
and covariance.  
These models have a different mean and covariance for each
market condition, but are regularized so that nearby market conditions
have similar models.
This technique allows us to fit models for market conditions that have
not occurred in the training data, by borrowing strength from
nearby market conditions for which we do have data.
These models are combined with a Markowitz-inspired optimization method
to yield a trading policy that is based on market conditions.
We illustrate our method on a small universe of 18 ETFs, using 
three well known and publicly available market variables to 
construct 1000 market conditions,
and show that it performs well out of sample.
The method, however, is general, and scales to much larger problems,
that presumably would use proprietary data sources and forecasts
along with publicly available data.
\end{abstract}

\section{Introduction}
\paragraph{Trading policy.}
We consider the problem of constructing a trading policy 
that depends on some observable market conditions,
as well as the current portfolio holdings.
We denote the asset daily returns as $y_t \in \reals^n$, 
for $t = 1, \ldots, T$.
The observable market conditions are denoted as $z_t$.
We assume these are discrete or categorical, so we have
$z_t \in \{1, \ldots, K\}$.
We denote the portfolio asset weights as $w_t \in \reals^n$,
with $\ones^T w_t=1$, where $\ones$ is the vector with all 
entries one.
The trading policy has the form
\[
\mathcal T : \{1,\ldots, K\} \times \reals^n \to \reals^n,
\]
where $w_t = \mathcal T(z_t,w_{t-1})$,
\ie, it maps the current market condition and previous portfolio
weights to the current portfolio weights.
In this paper we refer to $z_t$ as the market conditions,
since in our example it is derived from market conditions,
but in fact it could be anything known before the portfolio
weights are chosen, including proprietary forecasts or other data.
Our policy $\mathcal T$ is a simple Markowitz-inspired 
policy, based on
a Laplacian regularized stratified model of the asset return mean
and covariance; see, \eg, \cite{markowitz1952portfolio,grinold2000active,boyd2017multiperiod}.

\paragraph{Laplacian regularized stratified model.}
We model the asset returns, conditioned on market conditions, as 
Gaussian,
\[
y \mid z \sim \mathcal N (\mu_{z}, \Sigma_{z}),
\]
with 
$\mu_{z} \in \reals^n$ and $\Sigma_{z} \in \symm_{++}^n$ 
(the set of symmetric positive definite $n \times n$ matrices), 
$z = 1, \ldots, K$.   This is a stratified model, with 
stratification feature $z$.
We fit this stratifed model, \ie, 
determine the means $\mu_1, \ldots, \mu_K$ and 
covariances $\Sigma_1, \ldots, \Sigma_K$,
by minimizing the negative log-likelihood of historical training data,
plus a regularization term that encourages
nearby market conditions to have similar means and covariances.
This technique allows us to fit models for market conditions which have
not occurred in the training data, by borrowing strength from
nearby market conditions for which we do have data.
Laplacian regularized stratified models are discussed in, \eg,
\cite{danaher2014jointgraphicallasso,saegusa2016est,tuck2019mmdistlapl,
tuck2019stratmodels,tuck2020covstratmodels,tuck2021eigenstrat}.
One advantage of Laplacian regularized stratified models is they
are interpretable.  They are also auditable: we can easily check 
if the results are reasonable.

\paragraph{This paper.}
In this paper we present a single example of developing a 
trading policy as described above.  Our example is small,
with a universe of 18 ETFs, and we use market conditions
that are publicly available and well known.
Given the small universe and our use of widely available
market conditions, we cannot expect much in terms of 
performance, but we will see that the trading algorithm
performs well out of sample.
Our example is meant only as a simple illustration of the ideas;
the techniques we decribe can easily scale
to a universe of thousands of assets, and use
proprietary forecasts in the market conditions.

\paragraph{Outline.}
We start by reviewing Laplacian regularized models in \S\ref{s:lrsm}.
In \S\ref{s:data_set} we describe the data records and dataset we use.
In \S\ref{s:market_conditions} we describe the economic conditions 
with which we will stratify our return and risk models.
In \S\ref{s:return_model} and \ref{s:risk_model} we describe,
fit, and analyze the stratified return and risk models, respectively.
In \S\ref{s:port_con} we give the details of how our stratified return and risk 
models are used to create the trading policy $\mathcal T$.
We mention a few extensions and variations
of the methods in \S\ref{s:variations_extensions}.

\subsection{Related work}
A number of studies show that the underlying covariances of equities 
change during different market conditions,
such as when the market performs historically well or poorly (a 
``bull'' or ``bear'' market, respectively), or when there is historically
high or low volatility~\cite{erb1994stockcorrs,longin2001stockcorrs,ang2003regimeshift,
ang2004regimeshift,borland2012stockcorrs}. 
Modeling the dynamics of underlying statistical properties of assets is an area of
ongoing research.
Many model these statistical properties as occurring in hard regimes, and utilize 
methods such as hidden Markov models~\cite{ryden1998hmm,hastie2009elements,nystrup2018hmm}
or greedy Gaussian segmentation~\cite{hallac2016ggs} to model the transitions 
and breakpoints between the regimes.
In contrast, this paper assumes a hard regime model of our statistical
parameters, but our chief assumption is, informally speaking, that similar 
regimes have similar statistical parameters.

Asset allocation based on changing market conditions is a sensible
method for active portfolio management~\cite{ang2002regimeshift,ang2011regimechange,
nystrup2015hardregime,petre2015dynamicassetalloc}.
A popular method is to utilize convex optimization
control policies to dynamically allocate assets in a portfolio,
where the time-varying statistical properties are modeled as a hidden Markov 
model~\cite{nystrup2019drawdowncontrol}.

\section{Laplacian regularized stratified models}\label{s:lrsm}
In this section we review Laplacian regularized
stratified models, focussing on the specific models we will use;
for more detail see \cite{tuck2019stratmodels,tuck2020covstratmodels}.
We are given data records of the form $(z,y) \in \{1,\ldots, K\} 
\times \reals^n$,
where $z$ is the feature over which we stratify, and $y$ is the outcome.
We let $\theta \in \Theta$ denote the parameter values in our model.
The stratified model consists of a choice of parameter $\theta_z$ for 
each value of $z$.
In this paper, we construct two stratified models.  One is for return,
where $\theta_z\in \Theta= \reals^n$ is an estimate or forecast of return, and 
the other is for return covariance, where $\theta_z\in \Theta = \symm^n_{++}$ 
is the inverse covariance or precision matrix, and $\symm_{++}^n$
denotes the set of symmetric positive definite $n \times n$ matrices. 
(We use the precision matrix since it is the natural parameter in 
the exponential family representation of a Gaussian, and renders the fitting
problems convex.)

To choose the parameters $\theta_1,\ldots, \theta_K$,
we minimize
\BEQ
\sum_{k=1}^K \left( \ell_k(\theta_k) + r (\theta_k) \right) +
\mathcal L(\theta_1,\ldots, \theta_K).
\label{eq:strat-obj}
\EEQ
Here $\ell_k$ is the loss function, that depends on the training data $y_i$,
for $z_i = k$, typically a negative log-likelihood under our model 
for the data.
The function $r$ is the local regularizer, chosen to improve out
of sample performance of the model.

The last term in \eqref{eq:strat-obj} is the 
Laplacian regularization, which encourages
neighboring values of $z$, under some weighted graph, to have similar
parameters.
It is characterized by $W \in \symm^K$, a symmetric weight matrix with zero 
diagonal entries and nonnegative off-diagonal entries.
The Laplacian regularization has the form
\[
\mathcal L(\theta_1, \ldots, \theta_K)
= \frac{1}{2}\sum_{i,j=1}^K W_{ij} \|\theta_i - \theta_j\|^2,
\]
where the norm is the Euclidean or $\ell_2$ norm when $\theta_z$ is a
vector, and the Frobenius norm when $\theta_z$ is a matrix.
We think of $W$ as defining a weighted graph, with edges associated
with positive entries of $W$, with edge weight $W_{ij}$.
The larger $W_{ij}$ is, the more encouragement we give for 
$\theta_i$ and $\theta_j$ to be close.

When the loss and regularizer are convex,
the problem \eqref{eq:strat-obj} is convex, and so in principle is tractable~\cite{boyd2004convex}.
The distributed method introduced in \cite{tuck2019stratmodels}, 
which exploits the properties that the first two terms in the objective 
are separable across $k$, while the last term is separable across the 
entries of the parameters,
can easily solve very large instances of the problem.

A Laplacian regularized stratified model typically includes
several hyper-parameters, for example that scale the local 
regularization, or scale some of the entries in $W$.
We adjust these hyper-parameters by choosing some 
values, fitting the Laplacian regularized stratified model for each choice
of the hyper-parameters,
and evaluating the true loss function on a (held-out)
validation set. 
(The true loss function is often but not always the same as the 
loss function used in the fitting objective \eqref{eq:strat-obj}.)
We choose hyper-parameters that give the least, or nearly
least, true loss on the validation data, biasing our choice toward larger 
values, \ie, more regularization.

We make a few observations about Laplacian regularized stratified models.
First, they are interpretable, and we can check them for 
reasonableness by examining the values $\theta_z$, and how they
vary with $z$.  At the very least, we can examine the largest and smallest
values of each entry (or some function) of 
$\theta_z$ over $z \in \{1,\ldots, K\}$.

Second, we note that a Laplacian regularized stratified model can be created
even when we have no training data for some, or even many, values of $z$.
The parameter values for those values of $z$ are obtained 
by borrowing strength from their neighbors for which we do have data.
In fact, the parameter values for values of $z$ for which we have no data
are weighted averages of their neighbors.  This implies a number of interesting
properties, such as a maximum principle:  Any such value lies between
the minimum and maximum values of the parameter over those values of $z$
for which we have data.
\label{p-maximum-principal}

\section{Dataset}\label{s:data_set}
Our example considers $n=18$ ETFs as the universe of assets,
\BEAS
&\text{AGG, DBC, GLD, IBB, ITA, PBJ, TLT, VNQ, VTI,}\\
&\text{XLB, XLE, XLF, XLI, XLK, XLP, XLU, XLV, XLY}.
\EEAS
Each data record has the form $(y,z)$, where
$y \in \reals^{18}$ is the daily \emph{active} return
of each asset with respect to VTI,
from market close on the previous day until market close on that day,
and $z$ represents the market condition
known at the previous day's market close, described later in
\S\ref{s:market_conditions}.
(The daily active return of each asset with respect to VTI
is the daily return of that asset minus the daily return of VTI.)
Henceforth, when we refer to return or risk 
we mean active return or active risk, with respect to our benchmark VTI.
The benchmark VTI has zero active return and risk.

Our dataset spans 
March 2006 to December 2019, for a total of 3462 data points.
We first partition it into two subsets. The first, 
using data from 2006--2014, is used to fit the return
and risk models
as well as to choose the hyper-parameters in the 
return and risk models and the trading policy.
The second subset, with data in 2015--2019, is 
used to test the trading policy.
We then randomly partition the first subset into two parts: 
a training set consisting of 
80\% of the data records, and a validation set 
consisting of 20\% of the data records.
Thus we have three datasets: a training data set with 1780 data points
in the date range 2006--2014,
a validation set with 445 data points
also in the date range 2006--2014,
and a test dataset with 1237 data points in the date range 2015--2019.
We use 9 years of data to fit our models and choose hyper-parameters,
and 5 years of later data to test the trading policy.
The return data in the training and validation datasets
were winsorized (clipped) at their 1st and 99th percentiles.
The return data in the test dataset was not.

\section{Stratified market conditions}
\label{s:market_conditions}
Each data record also includes the market condition $z$ known on the 
previous day's market close.
To construct the market condition $z$,
we start with three (real-valued) market indicators.

\paragraph{Market implied volatility.}
The volatility of the market is a commonly used economic 
indicator, with extreme values associated with market 
turbulence~\cite{french1987vol,schwert1989vol,aggarwal1999vol,chun2020vol}.
Here, volatility is measured by the 15-day moving average of 
the CBOE volatility index (VIX) on 
the S\&P 500 \cite{VIX_site}, lagged by an additional day.


\paragraph{Inflation rate.}
The inflation rate measures the percentage change of purchasing power
in the economy~\cite{wynne1994inf,boyd1996inf,boyd2001inf,boyd2003inf,hung2003inf,mahyar2017inf}.
The inflation rate is published by the United States
Bureau of Labor Statistics~\cite{inflation_BLS} as the percent
change of the consumer price index (CPI), which measures changes
in the price level of a representative basket of consumer goods
and services, and is updated monthly.

\paragraph{30-year U.S. mortgage rates.}
This metric is the interest rate charged by a mortgage lender 
on 30-year mortgages, and the change of this rate is an economic 
indicator correlated with economic spending~\cite{lacava2016mort,sutton2017mort}.
The 30-year U.S. mortgage rate are published by the Federal Home 
Loan Mortgage Corporation, a public government-sponsored enterprise, 
and is generally updated weekly~\cite{mortgage30us}.

\bigskip

These three economic indicators are not particularly correlated 
over the training and validation period,
as can be seen in table~\ref{tab:market_cond_corrs}.
\begin{table}
\centering
\begin{tabular}{|l|c|c|c|c|}
\hline
  & Volatility & Inflation & Mortgage\\ \hline
  Volatility & 1 & -0.13 & -0.14\\
  Inflation & - & 1 & 0.21\\
  Mortgage & - & - & 1\\
\hline
\end{tabular}
\caption{Correlation of the market indicators 
over the training and validation period, 2006--2014.} 
\label{tab:market_cond_corrs}
\end{table}

\paragraph{Discretization.}
Each of these market indicators is binned into deciles, labeled $1, \ldots, 10$.
(The decile boundaries are computed using the data up to 2015.)
The total number of stratification feature values is then
$K=10\times 10 \times 10=1000$. 
We can think of $z$ as a 3-tuple of deciles, in $\{1,\ldots, 10\}^3$,
or encoded as a single value $z \in \{1,\ldots, 1000\}$.

\begin{figure}
\centering
\includegraphics[width=\textwidth]{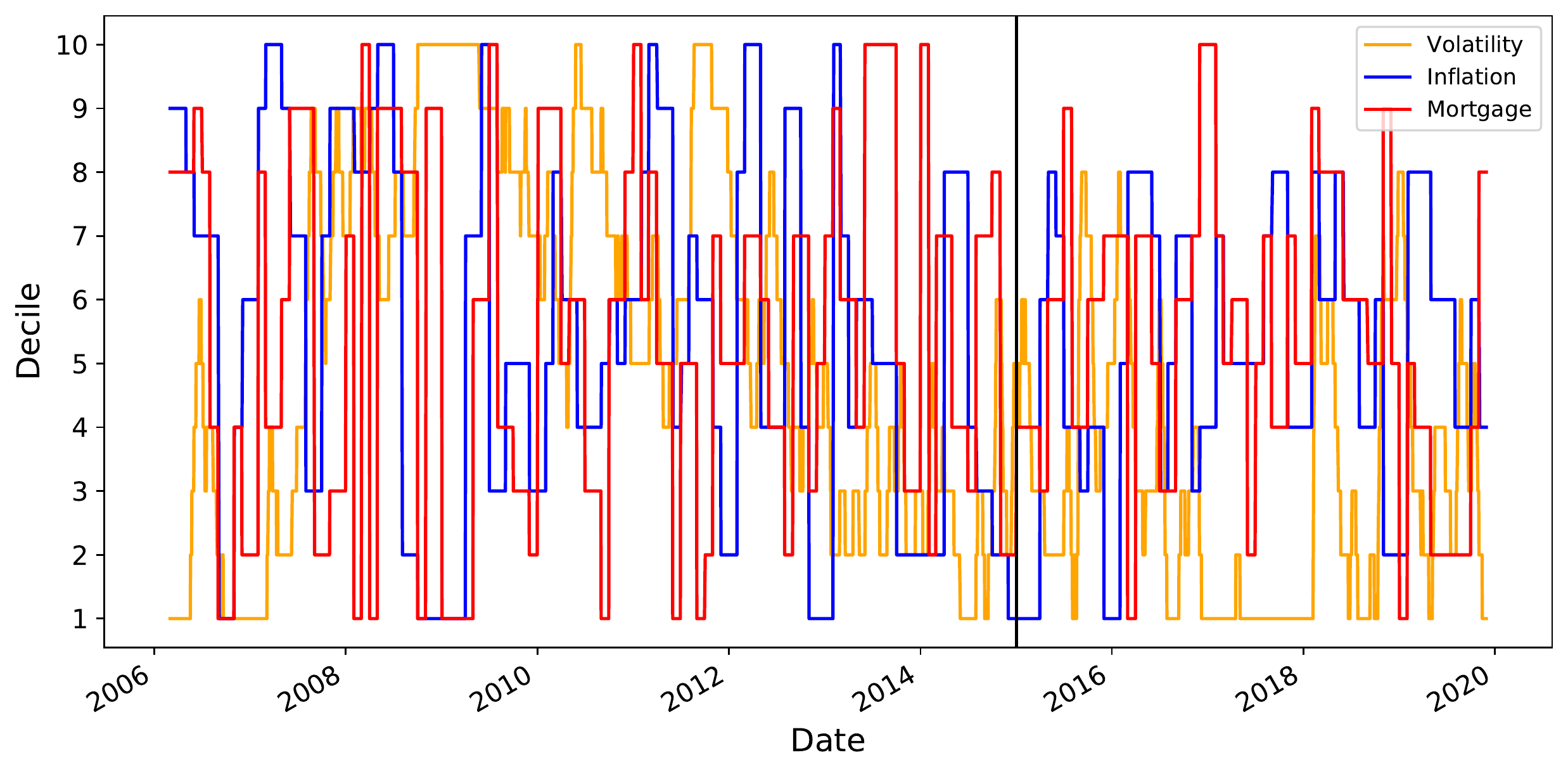}
\caption{Stratification feature values over time.  The vertical
line at 2015 separates the training and validation period
(2006--2014) from the test period (2015--2019).}
\label{fig:z_plot}
\end{figure}

The market conditions over the entire dataset are shown in
figure~\ref{fig:z_plot}, with the vertical line
at 2015 indicating the boundary between the training and validation
period (2006--2014) and the test period (2015--2019).
The average value of $\|z_{t+1}-z_t\|_1$ 
(interpreting them as vectors in $\{1,\ldots, 10\}^3$)
is around 0.26, meaning that on each day, the market conditions change by around 
0.26 deciles on average.

\paragraph{Data scarcity.}
The market conditions can take on $K=1000$ possible values.
In the training/validation datasets, only 224 of 1000 market conditions
appear, so there
are $776$ market conditions for which there are no data points.
The most populated market condition,
which corresponds to market conditions $(9,0,0)$, 
contains $84$ data points.
The average number of data points per market condition in the training/validation 
data is 2.23.

For over 77\% of the market conditions, we have \emph{no} training data.
This scarcity of data means that the Laplacian regularization is 
critical in constructing models of the return and risk that depend
on the market conditions.  

In the test dataset, only 133 of the economic conditions appear.
The average number of data points per market condition in the
test dataset is 1.26.
Only nine economic conditions appear in both the 
training/validation and test datasets.
In the test data, there are only 397 days 
(about 32\% of the 1237 test data days)
in which the market conditions 
for that day were observed in the training/validation datasets.

\paragraph{Regularization graph.}
Laplacian regularization requires a weighted graph that tells 
us which market conditions are `close'.
Our graph is the Cartesian product of three chain graphs, which 
link each decile of each indicator to its successor (and predecessor).
This graph on the 1000 values of $z$ has 2700 edges.
Each edge connects two adjacent deciles of one of our three economic 
indicators.
We assign three different positive weights to the edges, depending on
which indicator they link.
We denote these as
\BEQ\label{e-L-hyper-params}
\gamma_{\mathrm{vol}}, \quad
\gamma_{\mathrm{inf}}, \quad
\gamma_{\mathrm{mort}}.
\EEQ
These are hyper-parameters in our Laplacian regularization.
Each of the nonzero entries in the weight matrix $W$ is 
one of these values.   For example, the edge between
$(3,1,4)$ and $(3,2,4)$, which connects two values of $z$
that differ by one decile in Inflation, has weight 
$\gamma_{\mathrm{inf}}$.

\section{Stratified return model}\label{s:return_model}
In this section we describe the stratified return model.
The model consists of a mean return vector in $\mu_z \in \reals^{18}$ for each 
of $K=1000$ different market conditions, for a total of
$Kn = 18000$ parameters. 

The loss in \eqref{eq:strat-obj} is a Huber penalty,
\[
\ell_k(\mu_k) = \sum_{t: z_t = k} \ones^T H(\mu_k - y_t),
\]
where $H$ is the Huber penalty (applied entrywise above),
\BEAS
H(z) = \begin{cases}
z^2, & |z| \leq M\\
2M|z|-M^2, & |z| > M,
\end{cases}
\EEAS
where $M>0$ is the half-width, which we fix at the reasonable 
value $M=0.01$.
(This corresponds to the $79$th percentile of absolute return on
the training dataset.)
We use quadratic or $\ell_2$ squared local 
regularization in \eqref{eq:strat-obj},
\[
r(\mu_k) = \gamma_{\mathrm{ret,loc}}  \|\mu_k\|_2^2,
\]
where the positive regularization weight $\gamma_{\mathrm{ret,loc}}$ 
is another hyper-parameter.

The Laplacian regularization contains the three hyper-parameters
\eqref{e-L-hyper-params},
so overall our stratified return model has four hyper-parameters.

\subsection{Hyper-parameter search}\label{ss:hps_ret}
To choose the hyper-parameters for the stratified return model,
we start with a coarse grid search, evaluating all combinations of
\BEAS
\gamma_{\mathrm{ret,loc}} &=& 0.001, 0.01, 0.1,\\
\gamma_{\mathrm{vol}} &=& 1, 10, 100, 1000, 10000,\\
\gamma_{\mathrm{inf}} &=& 1, 10, 100, 1000, 10000,\\
\gamma_{\mathrm{mort}} &=& 1, 10, 100, 1000, 10000,
\EEAS
a total of $375$ combinations,
and selecting the hyper-parameter combination
that yielded the largest correlation between the 
return estimates and the returns over the validation set.
(Thus, our true loss is negative correlation of forecast 
and realized returns.)
The hyper-parameters
\[
(
\gamma_{\mathrm{ret,loc}}, 
\gamma_{\mathrm{vol}}, 
\gamma_{\mathrm{inf}}, 
\gamma_{\mathrm{mort}}) 
= 
(0.01, 10, 10, 1000)
\]
gave the best results over this coarse hyper-parameter grid search.

We then perform a finer hyper-parameter grid search, 
focusing on hyper-parameters around the best values from 
the coarse search.
We test all combinations of
\BEAS
\gamma_{\mathrm{ret,loc}} &=& 0.0075, 0.01, 0.0125,\\
\gamma_{\mathrm{vol}} &=& 2, 5, 10, 20, 50,\\
\gamma_{\mathrm{inf}} &=& 2, 5, 10, 20, 50,\\
\gamma_{\mathrm{mort}} &=& 200, 500, 1000, 2000, 5000,
\EEAS
a total of $375$ combinations.
The final hyper-parameter values are
\BEQ\label{e-return-final-hyps}
(
\gamma_{\mathrm{ret,loc}}, 
\gamma_{\mathrm{vol}}, 
\gamma_{\mathrm{inf}}, 
\gamma_{\mathrm{mort}}) 
=(0.0075, 10, 50, 5000).
\EEQ
These can be roughly interpreted as follows.  The large value 
for $\gamma_{\mathrm{mort}}$ tells us
that our return model should not vary much with mortage rate,
and the smaller values for
for $\gamma_{\mathrm{vol}}$ and $\gamma_{\mathrm{inf}}$ tells us that 
iour return model can vary more with volatility and inflation.

\subsection{Final stratifed return model}

Table~\ref{tab:results_return} shows the correlation coefficient of the 
return estimates to the true returns over the training and validation sets,
for the stratified return model and the common return
model, \ie, the empirical mean over the training set.
The stratified return model estimates have a larger correlation
with the realized returns in both the training set and the validation set.
The common return model even has a slightly negative correlation 
with the true returns on the validation dataset.

\begin{table}
\centering
\begin{tabular}{|l|c|c|c|c|}
\hline
Model & Train correlation & Validation correlation \\ \hline
Stratified return model & 0.097 & 0.052 \\
Common return model & 0.018 & -0.001 \\
\hline
\end{tabular}
\caption{Correlations to the true returns 
over the training set and the held-out validation set 
for the return models.}
\label{tab:results_return}
\end{table}

Table~\ref{tab:model_return_prediction_statistics} summarizes some 
of the statistics of our stratified return model 
over the 1000 market conditions, along with the common model value.
We can see that each forecast varies considerably across the market conditions.
Note that the common model values are the averages over the 
training data; the median, minimum, and maximum are over the 
1000 market conditions.

\begin{table}
 \centering
 \begin{tabular}{|l|c|c|c|c|}
  \hline
  Asset &  Common &  Median &    Min &    Max \\
  \hline
  AGG &  -0.021 &  -0.071 & -0.128 &  0.073 \\
  DBC &  -0.056 &  -0.060 & -0.158 &  0.106 \\
  GLD &  -0.012 &  -0.012 & -0.119 &  0.153 \\
  IBB &   0.033 &   0.031 & -0.098 &  0.139 \\
  ITA &   0.022 &   0.031 & -0.077 &  0.075 \\
  PBJ &   0.006 &   0.005 & -0.039 &  0.112 \\
  TLT &  -0.000 &  -0.063 & -0.173 &  0.110 \\
  VNQ &   0.016 &   0.009 & -0.301 &  0.071 \\
  XLB &   0.001 &   0.010 & -0.065 &  0.078 \\
  XLE &  -0.005 &   0.014 & -0.122 &  0.127 \\
  XLF &  -0.019 &  -0.040 & -0.398 &  0.055 \\
  XLI &   0.007 &   0.010 & -0.056 &  0.062 \\
  XLK &   0.005 &   0.004 & -0.059 &  0.090 \\
  XLP &   0.005 &  -0.004 & -0.041 &  0.070 \\
  XLU &  -0.008 &  -0.018 & -0.074 &  0.083 \\
  XLV &   0.010 &   0.009 & -0.033 &  0.065 \\
  XLY &   0.013 &   0.004 & -0.059 &  0.066 \\
  \hline
  \end{tabular}
 \caption{
Return predictions, in percent daily return.
The first column gives the common return model;
the second, third, and fourth columns
give median, minimum, and maximum return
predictions over the 1000 market conditions 
for the Laplacian regularized stratified model.}
 \label{tab:model_return_prediction_statistics}
\end{table}

\clearpage
\section{Stratified risk model}\label{s:risk_model}
In this section we describe the stratified risk model, \ie, 
a return covariance that depends on $z$.
For determining the risk model, we can safely ignore the
(small) mean return, and assume that $y_t$ has zero mean.
The model consists of $K=1000$ 
inverse covariance matrices $\Sigma_k^{-1} = \theta_k \in \symm_{++}^{18}$, 
indexed by the market conditions.
Our stratified risk model has $Kn(n+1)/2 = 171000$ parameters. 

The loss in \eqref{eq:strat-obj} 
is the negative log-likelihood on the training set
(scaled, with constant terms ignored),
\[
\ell_k(\theta_k) = 
\Tr(S_k \Sigma_{k}^{-1}) - \log\det(\Sigma_k^{-1})
\]
where $S_k = \frac{1}{n_k}\sum_{t : z_t = k} y_t y_t^T$ is the 
empirical covariance matrix of the data $y$ for which $z=k$,
and $n_k$ is the number of data samples with $z=k$.
(When $n_k=0$, we take $S_k=0$.)
We found that local regularization did not improve the model performance,
so we take local regularization $r = 0$.
All together our stratified risk model has the three Laplacian
hyper-parameters \eqref{e-L-hyper-params}.

\subsection{Hyper-parameter search}\label{ss:hps_risk}
We start with a coarse grid search over all $125$ combinations of
\BEAS
\gamma_{\mathrm{vol}} &=& 0.1, 1, 10, 100, 1000,\\
\gamma_{\mathrm{inf}} &=& 0.1, 1, 10, 100, 1000,\\
\gamma_{\mathrm{mort}} &=& 0.1, 1, 10, 100, 1000,
\EEAS
selecting the hyper-parameter combination with 
the smallest negative log-likelihood (our true loss)
on the validation set.
The hyper-parameters
\[
(\gamma_{\mathrm{vol}}, 
\gamma_{\mathrm{inf}}, 
\gamma_{\mathrm{mort}}) 
= (1, 1, 100)
\]
gave the best results.

We then perform a fine search, 
focusing on hyper-parameter value near
the best values from the coarse search.
We evaluate all $125$ combinations of
\BEAS
\gamma_{\mathrm{vol}} &=& 0.2, 0.5, 1, 2, 5,\\
\gamma_{\mathrm{inf}} &=& 0.2, 0.5, 1, 2, 5,\\
\gamma_{\mathrm{mort}} &=& 20, 50, 100, 200, 500.
\EEAS
For the stratified risk model, the final hyper-parameter 
values chosen are
\[
(\gamma_{\mathrm{vol}}, 
\gamma_{\mathrm{inf}}, 
\gamma_{\mathrm{mort}}) 
= (0.2, 5, 20).
\]
It is interesting to compare these to the hyper-parameter values
chosen for the stratified return model, 
given in \eqref{e-return-final-hyps}.  Since the losses
for return and risk models are different, we can scale the 
hyper-parameters in the return and risk to compare them.
We can see that they are not the same, but not too different,
either; both choose $\gamma_{\mathrm{inf}}$ larger than
$\gamma_{\mathrm{vol}}$, and
$\gamma_{\mathrm{mort}}$ quite a bit larger than
$\gamma_{\mathrm{vol}}$.

\subsection{Final stratified risk model}

Table~\ref{tab:results_risk} shows the average negative log likelihood 
(scaled, with constant terms ignored)
over the training and held-out validation sets, 
for both the stratified risk model and the common risk
model, \ie, the empirical covariance.
We can see that the stratified risk model has substantially better
loss on the training and validation sets.

\begin{table}
\centering
\begin{tabular}{|l|c|c|c|c|}
\hline
Model  & Train loss & Validation loss\\ \hline
Stratified risk model & -10.64 & -4.27 \\
Common risk model & 2.44 & 3.26 \\
\hline
\end{tabular}
\caption{Average negative log-likelihood (scaled, with constant terms ignored)
over the training and validation 
sets for the stratified and common risk models.}
\label{tab:results_risk}
\end{table}

Table~\ref{tab:model_risk_prediction_statistics} summarizes some 
of the statistics of our stratified return model asset 
volatilities, \ie, $\left((\Sigma_z)_{ii}\right)^{1/2}$, 
expressed as daily percentages,
over the 1000 market conditions, along with the common model 
asset volatilities.
We can see that the predictions vary considerably across the 
market conditions, with a few varying by a factor almost up to ten.
Table~\ref{tab:model_correl_prediction_statistics} summarizes the same
statistics for the correlation of each asset with AGG, an aggregate bond market 
ETF.  Here we see dramatic variation,
for example, the correlation between XLI (an industrials ETF) 
and AGG varies from
-85\% to +80\% over the market conditions.

\begin{table}
 \centering
 \begin{tabular}{|l|c|c|c|c|}
    \hline
    Asset &  Common &  Median &    Min &    Max \\
    \hline
    AGG &   1.313 &   0.864 &  0.537 &  4.236 \\
    DBC &   1.289 &   0.998 &  0.725 &  3.950 \\
    GLD &   1.665 &   1.194 &  0.866 &  5.613 \\
    IBB &   0.914 &   0.796 &  0.634 &  1.920 \\
    ITA &   0.619 &   0.549 &  0.474 &  1.421 \\
    PBJ &   0.648 &   0.502 &  0.414 &  1.502 \\
    TLT &   1.816 &   1.263 &  0.734 &  6.050 \\
    VNQ &   1.328 &   0.769 &  0.643 &  3.730 \\
    XLB &   0.772 &   0.623 &  0.491 &  2.148 \\
    XLE &   1.024 &   0.793 &  0.628 &  3.117 \\
    XLF &   1.190 &   0.602 &  0.378 &  3.479 \\
    XLI &   0.499 &   0.432 &  0.360 &  1.008 \\
    XLK &   0.515 &   0.453 &  0.380 &  1.241 \\
    XLP &   0.760 &   0.569 &  0.424 &  1.682 \\
    XLU &   0.883 &   0.724 &  0.614 &  2.155 \\
    XLV &   0.703 &   0.499 &  0.417 &  1.560 \\
    XLY &   0.536 &   0.433 &  0.350 &  1.386 \\
    \hline
  \end{tabular}
 \caption{
Forecasts of volatility, expressed in percent daily 
return.
The first column gives the common model;
the second, third, and fourth columns
give median, minimum, and maximum volatility
predictions over the 1000 market conditions 
for the Laplacian regularized stratified model.}
 \label{tab:model_risk_prediction_statistics}
\end{table}

\begin{table}
 \centering
 \begin{tabular}{|l|c|c|c|c|}
  \hline
  Asset &  Common &  Median &    min &    max \\
  \hline
  AGG &   1 &   1 &  1 &  1 \\
  DBC &   0.490 &   0.414 & -0.285 &  0.959 \\
  GLD &   0.683 &   0.522 & -0.131 &  0.979 \\
  IBB &   0.238 &   0.066 & -0.669 &  0.888 \\
  ITA &   0.021 &  -0.059 & -0.896 &  0.842 \\
  PBJ &   0.569 &   0.356 & -0.058 &  0.918 \\
  TLT &   0.934 &   0.888 &  0.749 &  0.995 \\
  VNQ &  -0.345 &   0.007 & -0.908 &  0.796 \\
  XLB &  -0.213 &  -0.216 & -0.802 &  0.826 \\
  XLE &  -0.203 &  -0.166 & -0.832 &  0.854 \\
  XLF &  -0.520 &  -0.267 & -0.946 &  0.105 \\
  XLI &  -0.108 &  -0.117 & -0.848 &  0.801 \\
  XLK &   0.158 &   0.091 & -0.749 &  0.864 \\
  XLP &   0.717 &   0.561 &  0.228 &  0.938 \\
  XLU &   0.555 &   0.427 &  0.010 &  0.945 \\
  XLV &   0.600 &   0.390 & -0.275 &  0.917 \\
  XLY &  -0.059 &  -0.035 & -0.833 &  0.534 \\
  \hline
\end{tabular}
 \caption{
Forecasts of correlations with the aggregate bond index AGG.
The first column gives the common model;
the second, third, and fourth columns
give median, minimum, and maximum correlation
predictions over the 1000 market conditions 
for the Laplacian regularized stratified model.}
 \label{tab:model_correl_prediction_statistics}
\end{table}

\clearpage
\section{Trading policy and backtest}\label{s:port_con}

\subsection{Trading policy}
In this section we give the details of how we use our stratified return 
and risk models to construct the trading policy $\mathcal T$.

At the beginning of each day $t$, we use the previous day's market conditions
$z_t$ to allocate our current portfolio according to the weights $w_t$,
computed as the solution of the Markowitz-inspired problem 
\cite{boyd2017multiperiod}
\BEQ
\begin{array}{ll}
\mbox{maximize} & 
\mu_{z_t}^T w
- \gamma_\mathrm{sc} \kappa^T (w)_{-}
- \gamma_\mathrm{tc} \tau_t^T |w-w_{t-1}|\\
\mbox{subject to} &
             w^T \Sigma_{z_t} w \leq \sigma^2, \quad 
             \ones^T w = 1,\\
             & \|w\|_1 \leq L_{\mathrm{max}}, \quad
             w_{\mathrm{min}}\leq w \leq w_{\mathrm{max}},
\end{array}
\label{eq:port_con}
\EEQ
with optimization variable $w \in \reals^{18}$, where 
$w_- = \max\{0,-w\}$ (elementwise), and the absolute value is
elementwise.
We describe each term and constraint below.
\BIT
\item \emph{Return forecast.}
The first term in the objective, $\mu_{z_t}^T w$, 
is the expected return under our forecast
mean, which depends on the current market conditions.
\item \emph{Shorting cost.} 
The second term 
$\gamma_\mathrm{sc} \kappa^T (w)_{-}$
is a shorting cost,
with $\kappa\in\reals_{+}^{18}$ the vector of shorting cost rates.
(For simplicity we take the shorting cost rates as constant.)
The positive hyper-parameter $\gamma_\mathrm{sc}$ 
scales the shorting cost term, and is used to control our shorting aversion.
\item \emph{Transaction cost.}
The third term $\gamma_\mathrm{tc} \tau_t^T |w-w_{t-1}|$ is a transaction cost, with
$\tau_t\in\reals_+^{18}$ the vector of transaction cost rates used 
on day $t$.
We take $\tau_t$ as one-half
the average bid-ask spread of each asset for the previous 15 trading days
(excluding the current day).
We summarize the bid-ask spreads of each asset over the training and holdout periods
in table~\ref{tab:tau}.
The positive hyper-parameter $\gamma_\mathrm{tc}$ 
scales the transaction cost term, and is used to control the turnover.
\item \emph{Risk limit.}
The constraint $w^T \Sigma_z w \leq \sigma^2$ limits the (daily)
risk (under our risk model, which depends on market conditions) to $\sigma$, 
which corresponds to an annualized risk of $\sqrt{250} \sigma$.
\item \emph{Leverage limit.}
The constraint $\|w\|_1 \leq L_{\mathrm{max}}$ limits the portfolio leverage,
or equivalently, it limits the total short position $\ones^T (w)_-$ to no more
than $(L-1)/2$.
\item \emph{Position limits.}
The constraint $w_\mathrm{min} \leq w \leq w_\mathrm{max}$ 
(interpeted elementwise) limits the individual weights.  
\EIT

\paragraph{Parameters.}
Some of the constants in the trading policy \eqref{eq:port_con} 
we simply fix to reasonable values.
We fix the shorting cost rate vector to $(0.0005)\ones$, \ie, 5 basis points
for each asset.
We take $\sigma = 0.0045$, which corresponds to an annualized
volatility of $\sqrt{250}\sigma$, around 7.1\%.
We take $L_\mathrm{max} = 2$, which means the total short position
cannot exceed one half of the portfolio value.
(A portfolio with a leverage of 2 satisfying $1^T w = 1$
is commonly referred to as a \emph{150/50 portfolio}.)
We fix the position limits as $w_\mathrm{min} = -0.25 \ones$ and
$w_\mathrm{max} = 0.4 \ones$, meaning we cannot short any asset by
more than $0.25$ times the portfolio value, and we cannot hold more than
$0.4$ times the portfolio value of any asset.

\begin{table}
 \centering
\begin{tabular}{|l|c|c|}
\hline
Asset  & Training/validation period & Holdout period \\
\hline
AGG    &  0.000298 &  0.000051 \\
DBC    &  0.000653 &  0.000324 \\
GLD    &  0.000112 &  0.000048 \\
IBB    &  0.000418 &  0.000181 \\
ITA    &  0.000562 &  0.000175 \\
PBJ    &  0.000966 &  0.000637 \\
TLT    &  0.000157 &  0.000048 \\
VNQ    &  0.000394 &  0.000066 \\
VTI    &  0.000204 &  0.000048 \\
XLB    &  0.000310 &  0.000098 \\
XLE    &  0.000181 &  0.000077 \\
XLF    &  0.000359 &  0.000200 \\
XLI    &  0.000295 &  0.000079 \\
XLK    &  0.000324 &  0.000093 \\
XLP    &  0.000298 &  0.000095 \\
XLU    &  0.000276 &  0.000099 \\
XLV    &  0.000271 &  0.000070 \\
XLY    &  0.000334 &  0.000059 \\
\hline
\end{tabular}
 \caption{One-half the mean bid-ask spread
 of each asset, over the training and validation periods and the holdout period.}
 \label{tab:tau}
\end{table}

\paragraph{Hyper-parameters.}
Our trading policy has two hyper-parameters,
$\gamma_\mathrm{sc}$ and
$\gamma_\mathrm{tc}$, which control our aversion to shorting and trading,
respectively.

\subsection{Backtests}
Backtests are carried out
starting from a portfolio of all VTI
and a starting portfolio value of $v=1$.
On day $t$, after computing $w_t$ as the solution 
to~\eqref{eq:port_con}, we compute the value of our portfolio
$v_t$ by
\[
r_{t, \mathrm{net}} = r_t^T w_t 
- \kappa^T (w_t)_{-} 
- (\tau_t^{\mathrm{sim}})^T|w_t-w_{t-1}|, \qquad
v_t = v_{t-1} (1+r_{t, \mathrm{net}}), 
\]
Here 
$r_t \in \reals^{18}$ is the vector of asset returns on day $t$,
$r_t^T w_t$ is the gross return of the portfolio for day $t$,
$\tau_t^{\mathrm{sim}}$ is one-half the realized bid-ask spread on day $t$,
and $r_{t, \mathrm{net}}$ is the net return of the portfolio for day $t$
including shorting and transaction costs.
In particular, 
\emph{our backtests take shorting and transaction costs into account}.
Note also that in the backtests, we use the actual realized
bid-ask spread on that day (which is not known at the beginning
of the day) to determine the true transaction cost,
whereas in the policy, we use the trailing 15 day average (which is
known at the beginning of the day).

Our backtest is a bit simplified.
Our simulation assumes dividend reinvestment.  We account for the 
shorting and transaction costs by adjusting the portfolio return,
which is equivalent to splitting these costs across the whole portfolio;
a more careful treatment might include a small cash account.
For portfolios of very high value, we would add an additional nonlinear
transaction cost term, for example proportional to the $3/2$-power
of $|w_t-w_{t-1}|$ \cite{boyd2017multiperiod}.

\subsection{Hyper-parameter selection}
To choose values of the two hyper-parameters in the trading policy,
we carry out multiple backtest simulations over 
the training set.
We evaluate these backtest simulations by their realized return 
(net, including costs) over the validation set.

\begin{figure}
 \centering
 \includegraphics[width=\textwidth]{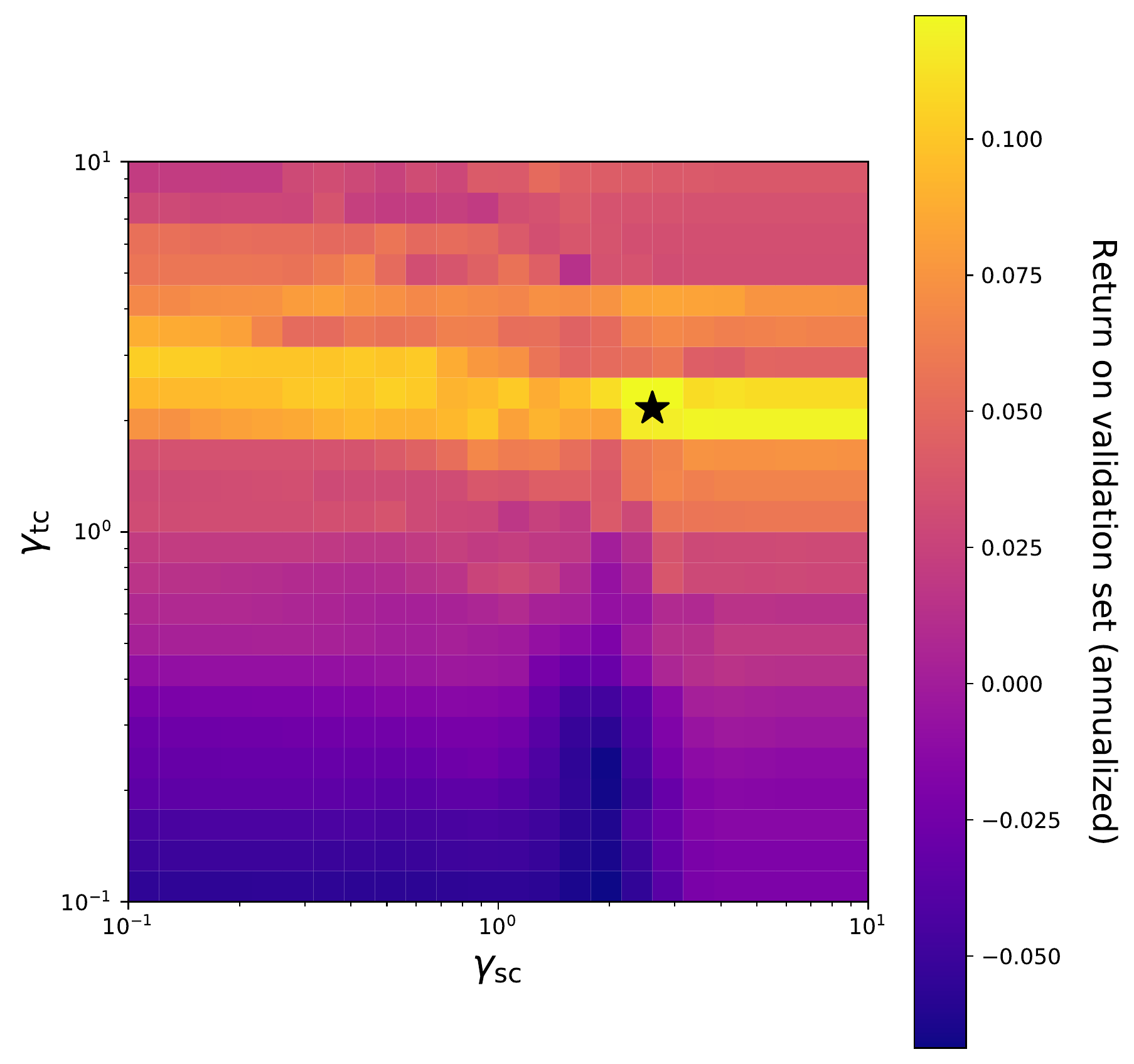}
 \caption{Heatmap of the annualized return on the validation set 
 as a function of the 
 two hyper-parameters $\gamma_\mathrm{sc}$ and $\gamma_\mathrm{tc}$.
 The star shows the hyper-parameter combination used in our trading policy.}
 \label{fig:portcon_grid_search}
\end{figure}

We perform a grid search, testing all 625 pairs of 25 values of each
hyper-parameter logarithmically spaced from $0.1$ to $10$.
The annualized return on the validation set, as a function of 
the hyper-parameters,
are shown in figure~\ref{fig:portcon_grid_search}.
We choose the final values 
\[
\gamma_\mathrm{sc} = 2.61, \quad
\gamma_\mathrm{tc} = 2.15,
\]
shown on figure~\ref{fig:portcon_grid_search} as a star.

These values are themselves interesting.  Roughly speaking,
we should plan our trades as if the shorting cost were more than 2.5 times
the actual cost, and the transaction cost is more than double the
true transaction cost.  
The blue and purple
region at the bottom of the heat map indicates poor validation performance
when the transaction cost parameter is too low, \ie, the policy
trades too much.

Table~\ref{tab:policy_train_test_results} gives the annualized return and
risk 
for the policies over the train and validation periods.
\begin{table}
\centering
\begin{tabular}{|l|c|c|}
\hline
& Return & Risk \\ \hline
Train & 11.2\%& 5.89\% \\
Validation & 12.2\% & 6.23\%\\
\hline
\end{tabular}
\caption{Annualized return and risk for the stratified
model policy over the train and validation periods.}
\label{tab:policy_train_test_results}
\end{table}

\paragraph{Common model trading policy.}
We will compare our stratified model trading policy 
to a common model trading policy,
which uses the constant return and risk models, along with
the same Markowitz policy \eqref{eq:port_con}.
In this case, none of the parameters in the optimization problem
change with market conditions, and the only parameter that changes
in different days is $w_{t-1}$, the previous day's asset weights,
which enters into the transaction cost.

We also perform a grid search for this trading policy, over
the same $625$ pairs of the hyper-parameters.
For the common model trading policy,
we choose the final values 
\[
\gamma_\mathrm{sc} = 0.12, \quad
\gamma_\mathrm{tc} = 1.47.
\]

\subsection{Final trading policy results}
We re-fit our stratified risk and return models, utilizing all of the data
in the training and validation sets, using the hyper-parameters selected 
in \S\ref{ss:hps_ret} and \S\ref{ss:hps_risk}.
We backtest our trading policy on the test dataset, which
includes data from 2015--2019. 
We remind the reader that no data from this date range
was used to create, tune, or validate any of the models, or to 
choose any hyper-parameters.
For comparison, we also give results of a backtest using
the constant return and risk models.

\begin{figure}
\centering
\includegraphics[width=\textwidth]{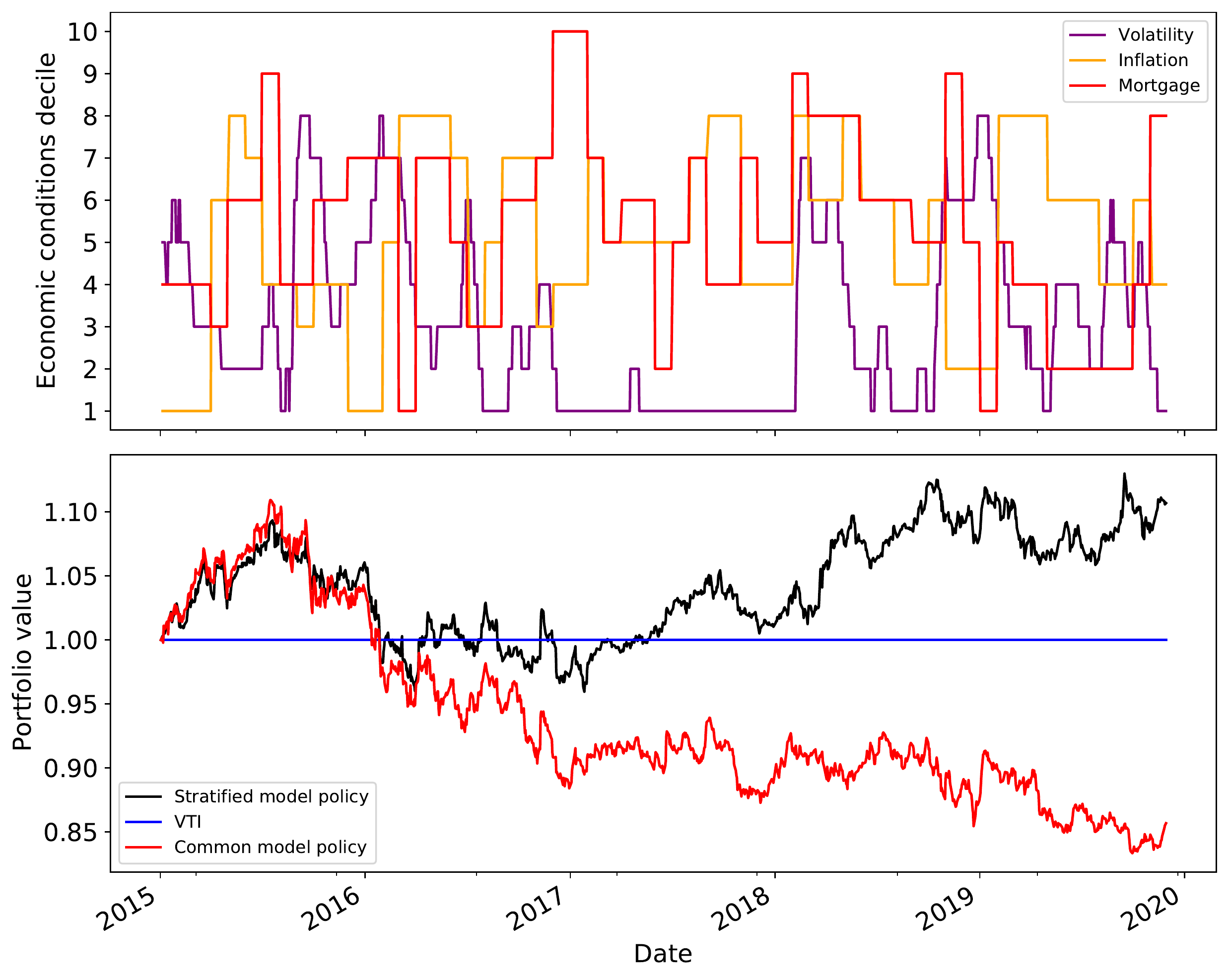}
\caption{Plot of economic conditions (top) and cumulative portfolio
value for the stratified model and 
the common model (bottom) over the test period. The horizontal
blue line is the cumulative portfolio value for buying and 
holding the benchmark VTI.}
\label{fig:policy}
\end{figure}

Figure~\ref{fig:policy} plots 
the economic conditions over the test period (top)
as well as the active portfolio value (\ie, value above the 
benchmark VTI)
for our stratified model and common model.
Buying and holding the benchmark VTI gives zero active return,
and a constant active portfolio value of 1.
The superior performance of the stratified model policy, 
\eg, higher return and lower volatility, is evident in this plot.

Table~\ref{tab:policy_annualized_results} shows 
the annualized active return, annualized active risk,
annualized active Sharpe ratio (return divided by risk), 
and maximum drawdown of the active portfolio value
for the policies over the test period.
We remind the reader that we are fully accounting for the 
shorting and transaction cost,
so the turnover of the policy is accounted for in these backtest metrics.
\begin{table}
\centering
\begin{tabular}{|l|c|c|c|c|c|}
\hline
  & Return & Risk & Sharpe ratio & Maximum drawdown \\ \hline
Stratified model policy & 2.30\% & 7.00\% & 0.33 & 12.2\%\\
Common model policy & -2.82\% & 7.73\% & -0.366 & 24.9\% \\
\hline
\end{tabular}
\caption{Annualized active return, active risk, active Sharpe ratios, 
and maximum drawdown of the active portfolio value 
for the three policies over the test 
period (2015--2019).}
\label{tab:policy_annualized_results}
\end{table}

The results are impressive when viewed in the following light.
First, we are using a very small universe of only 18 ETFs.
Second, our trading policy uses only three widely 
available market conditions, and indeed, only their deciles.
Third, the policy was entirely developed using data prior to 
2015, with no adjustments made for the next five years.
(In actual use, one would likely re-train the model periodically,
perhaps every quarter or year.)

\begin{figure}
\centering
\includegraphics[width=\textwidth]{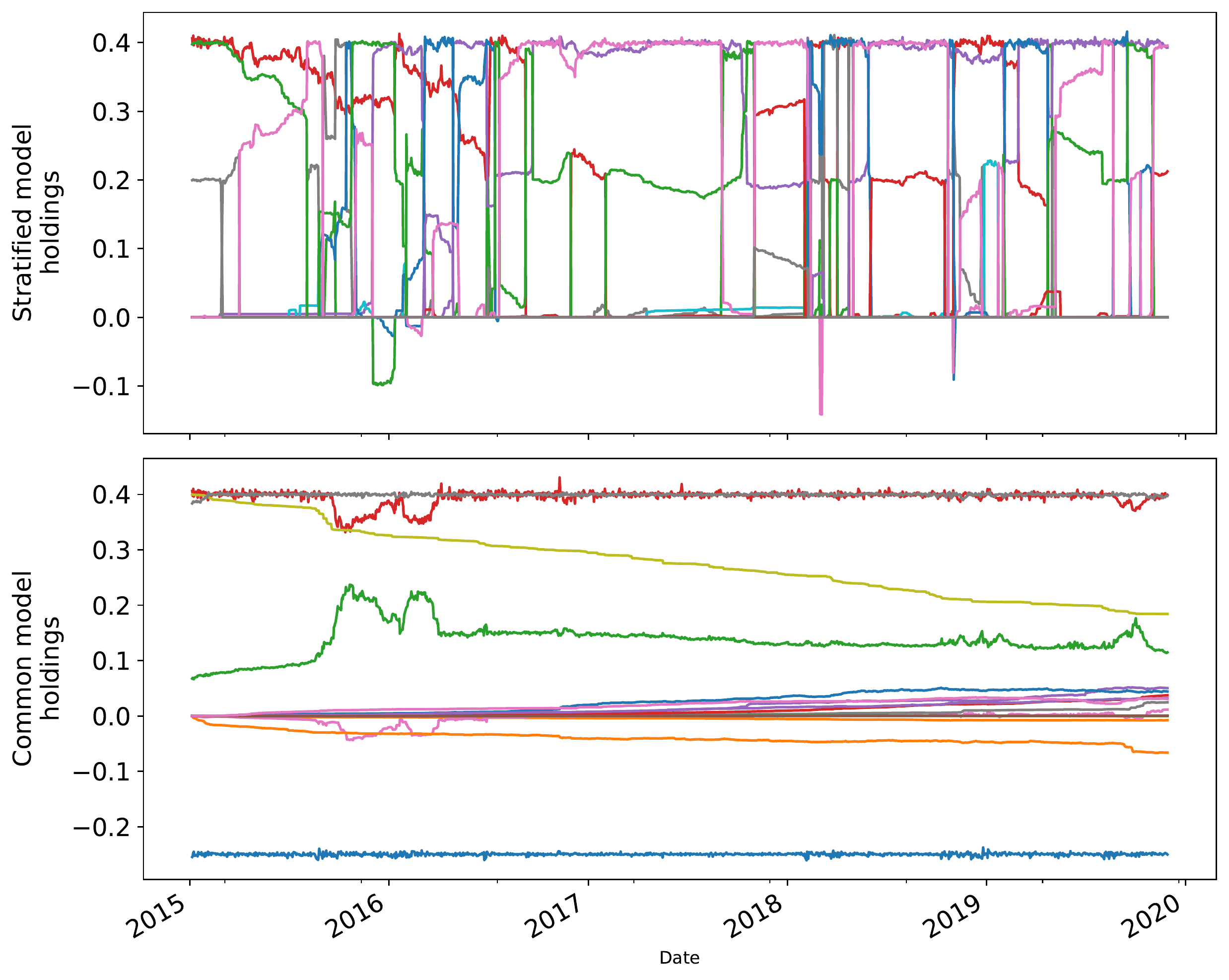}
\caption{Asset weights of the stratified model policy (top) and of 
the common model policy (bottom), over the test period.
The first time period asset weights, 
which are all VTI, are not shown.}
\label{fig:holdings}
\end{figure}

\paragraph{Comparison of stratified and constant policies.}
In figure~\ref{fig:holdings}, we plot the asset weights 
of the stratified model policy (top) and of 
the common model policy (bottom), over the test period.
(The variations in the common model policy holdings come from a combination of
a daily rebalancing of the assets and the transaction cost model.)
The top plot shows that the weights in the stratified policy change 
considerably with market conditions.  
The only assets that are shorted to a significant degree are 
AGG, GLD and TLT, and only during times of market turbulence.
The common model policy is mainly concentrated in just seven assets,
AGG (bonds) 
GLD (gold),
IBB (biotech),
XLE (energy),
XLF (financials),
XLY (consumer discretionary),
and VTI 
(which is effectively cash when considering active returns and risks),
and never shorts any assets, \ie, is long only.
Moreover, the common model policy shorts AGG (to the position limit of -0.25) 
and TLT and XLF (by a much lesser degree).

\paragraph{Factor analysis.}
We fit a linear regression model of the active 
returns of the two policies 
over the test set to four of the 
Fama-French 
factors~\cite{fama1992famafrenchfactors,fama1993famafrenchfactors,french2021famafrench}:
\begin{itemize}
 \item \emph{MKTRF}, the value-weighted return of 
 United States equities, minus the risk free rate,
 \item \emph{SMB}, the return on a portfolio of small size stocks 
 minus a portfolio of big size stocks,
 \item \emph{HML}, the return on a portfolio of value stocks 
 minus a portfolio of growth stocks, and
 \item \emph{UMD}, the return on a portfolio of high momentum stocks
 minus a portfolio of low or negative momentum stocks.
\end{itemize}
We also include an intercept term, commonly referred to as alpha.
Table~\ref{tab:fama_french} gives the results of these fits.
Relative to the common model policy, the stratified model policy active returns
are much less positively correlated to the market,
shorter the size factor,
longer the value factor,
and shorter the momentum factor.
Its active alpha is around 2.43\% annualized.
While not very impressive on its own, this alpha seems good considering
it was accomplished with just 18 ETFs, and using only three 
widely available quantities in the policy.

\begin{table}
 \centering
 \begin{tabular}{|l|c|c|c|}
  \hline
  Factor &  Stratified model policy &  Common model policy \\
  \hline
  MKTRF &                -0.033887 &             0.154132 \\
  SMB   &                 0.165571 &             0.231877 \\
  HML   &                -0.233028 &            -0.457454 \\
  UMD   &                -0.127748 &            -0.108726 \\
  Alpha &                 0.000097 &            -0.000228 \\
  \hline
    \end{tabular}   
 \caption{The top four rows give the regression model coefficients of the 
active portfolio returns on the Fama-French factors; the fifth row gives the
intercept or alpha value.}
 \label{tab:fama_french}
\end{table}

\section{Extensions and variations}\label{s:variations_extensions}

We have presented a simple (but realistic) 
example only to illustrate the ideas, which can
easily be applied in more complex settings, with a far larger
universe, a more complex trading policy, 
and using proprietary forecasts of returns and quantities 
used to judge market conditions.  
We describe some extensions and variations on our method below.

\paragraph{Multi-period optimization.}
For simplicity we use a policy that is based on solving 
a single-period Markowitz problem.  The entire method immediately
extends to policies based on multi-period optimization.
For example, we would fit separate stratified models of return and risk
for the next 1 day, 5 day, 20 day, and 60 day periods (roughly
daily, weekly, monthly, quarterly), all based on the same
current market conditions.
These data are fed into a multi-period optimizer as described in 
\cite{boyd2017multiperiod}.

\paragraph{Joint modeling of return and risk.}
In this paper we created separate Laplacian regularized 
stratified models for return and risk.  The advantage of this approach
is that we can judge each model separately (and with different true objectives),
and use different hyper-parameter values.
It is also possible to fit the return mean and covariance \emph{jointly},
in one stratified model, using the natural parameters in the 
exponential family for a Gaussian, $\Sigma^{-1}$ and $\Sigma^{-1}\mu$.
The resulting log-likelihood is jointly concave, and a Laplacian regularized
model can be directly fit.

\paragraph{Low-dimensional economic factors.}
When just a handful (such as in our example, three) base quantities are
used to construct the stratified market conditions,
we can bin and grid the values as we do in this paper.
This simple stratification of market conditions 
preserves interpretability.
If we wish to include more raw data in our stratification of market
conditions, simple binning and enumeration is not practical.
Instead we can use several techniques to handle such situations.
The simplest is to perform dimensionality-reduction
on the (higher-dimensional) economic conditions, such as principal component 
analysis~\cite{pearson1901pca} or low-rank
forecasting~\cite{barratt2020lowrank}, and appropriately bin these 
low-dimensional economic conditions.
These economic conditions may then be related on a graph with edge weights
decided by an appropriate method, such as nearest neighbor weights.

\paragraph{Structured covariance estimation.}
It is quite common to model the covariance matrix of returns as having
structure, \eg, as the sum of a diagonal matrix plus a low rank 
matrix~\cite{richard2012covmtx,fan2016covmtx}.
This structure can be added by a combination of introducing new variables 
to the model and encoding constraints in the local regularization.
In many cases, this structure constraint turns the stratified risk
model fitting problem into a non-convex problem, which may be 
solved approximately.

\paragraph{Multi-linear interpolation.}
In the approach presented above, the economic conditions are categorical,
\ie, take on one of $K=1000$ possible values at each time $t$, based on
the deciles of three quantities.
A simple extension is to use multi-linear interpolation 
\cite{weiser1988interp,davies1997interp} to determine the return
and risk to use in the Markowitz optimizer.
Thus we would use the actual quantile of the three market quantiities,
and not just their deciles.
In the case of risk, we would apply the interpolation to
the precision matrix $\Sigma_t^{-1}$, the natural parameter in 
the exponential family description of a Gaussian.

\paragraph{End-to-end hyper-parameter optimization.}
In the example presented in this paper there are a total 
of nine hyper-parameters to select.
We keep things simple by separately optimizing the hyper-parameters 
for the stratified return model, the stratified risk model, and 
the trading policy. This approach allows each step to be checked independently.
It is also possible to simultaneously optimize all of the hyper-parameters
with respect to a single backtest, using, for example,
CVXPYlayers \cite{agrawal2019cvxpylayers,agrawal2019cocp} to differentiate through the trading policy.

\paragraph{Stratified ensembling.}
The methods described in this paper can be used to combine or emsemble 
a collection of different return forecasts or signals, whone performance
varies with market (or other) conditions.
We start with a collection of return predictions, and combine these
(ensemble them) using weights that are a function of the market conditions.
We develop a stratified selection of the combining weights.

\section{Conclusions}

We argue that stratified models are interesting and useful in portfolio
construction and finance. 
They can contain a large number of parameters,
but unlike, say, neural networks, they are fully interpretable and auditable.
They allow arbitrary variation across market conditions, with Laplacian 
regularization there to help us come up with reasonable models
even for market conditions for which we have no training data.
The maximum principle mentioned on page \pageref{p-maximum-principal} tells us that
a Laplacian regularized stratified model will never do anything crazy when it 
encounters values of $z$ that never appeared in the training data.
Instead it will use a weighted sum of other values for which we do have
training data.  These weights are not just any weights, but ones 
carefully chosen by validation.

The small but realistic example we have presented is only meant to 
illustrate the ideas.  The very same ideas and method 
can be applied in far more complex and 
sophisticated settings, with a larger universe of assets, a more 
complex trading policy, and incorporating proprietary data and forecasts.

\section*{Acknowledgements}
The authors gratefully acknowledge discussions with and suggestions from Ronald Kahn,
Raffaele Savi, and Andrew Ang.
Jonathan Tuck is supported by the Stanford Graduate 
Fellowship in Science and Engineering.

\clearpage
\bibliography{refs}

\end{document}